\documentclass[preprint]{elsarticle}
\usepackage{graphics}
\usepackage{graphicx}
\usepackage{amssymb}
\newcommand{\beq}{\begin{equation}}
\newcommand{\eeq}{\end{equation}}
\newcommand{\ber}{\begin{eqnarray}}
\newcommand{\eer}{\end{eqnarray}}
\begin{document}
\begin{frontmatter}
\title{Competing order parameters and a tricritical point with a difference}
\author{Arghya Dutta \corref{cor}}
\ead{arghya@bose.res.in}
\author{J. K. Bhattacharjee}
\address{S. N. Bose National Centre for Basic Sciences\\ Block-JD, Sector-3, Salt Lake, Kolkata, Pin-700098, India.}
\cortext[cor]{Corresponding author. Tel: +91 (033) 2335 5706 fax:
+91 (033) 2335 3477}

\begin{abstract}
We propose a mean-field, phenomenological Ginzburg-Landau free energy functional with two competing order parameters for a two-component, spin-polarized Fermi gas. This free energy supports a tricritical point which is different from the conventional one and this change offers a correct understanding of the experimental phase diagram of imbalanced Fermi systems ( Shin et al, Nature, 2008). The specific heat also happens to be different than in standard theory.
\end{abstract}
\end{frontmatter}
%
\section{Introduction}
Population imbalance between spin-up and spin-down fermions in a mixture of two Fermi gases leads to exotic superfluid phases. In past ten years, these exotic superfluid phases have been extensively studied, both theoretically\cite{liu03,forbes05,machida06,hu06,pilati08} and experimentally\cite{patridge06,zwierlein06,zwierlein06nat,shin06,schunck07,patridge06prl} in ultra-cold atomic gases  and in quark-gluon plasma which resides in the core of neutron stars\cite{rajagopal01,casal04}. Similar imbalanced fermionic systems were studied in electron superconductors in a magnetic field as early as nineteen-sixties\cite{sarma63,maki64}, and showed the possibility of tricritical points where the second-order and the first-order lines meet along with the line separating stable and unstable superconducting phases. That the tricritical point is fundamental to the understanding of superfluidity of polarized, two-component Fermi gases was first pointed out by Parish et al\cite{parish07}.

\begin{figure}
\centering
\includegraphics[scale=0.5]{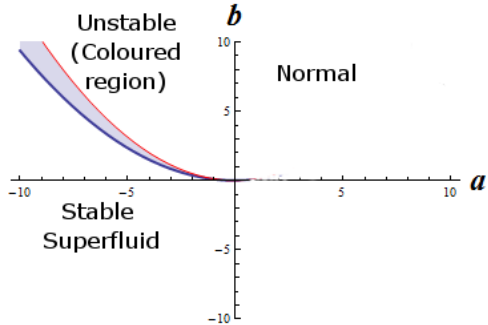}
\caption{Standard $\mathcal{O}(\psi^6)$ theory phase diagram as a function of the scaled temperature ($a$) and temperature-independent, polarization-dependent fourth order coupling constant($b$). The shaded region represents a thermodynamically-unstable, finite-$\psi$ region.}
\label{psi}
\end{figure}

The standard model for studying tricritical point starts with the following Ginzburg-Landau free energy per unit volume: $F=\frac{a}{2}\psi^2+\frac{b}{4}\psi^4+\frac{c}{6}\psi^6$, where $\psi$ is superfluid order parameter, $a$ is scaled temperature, $c$ is a positive constant, and $b$ can take both positive and negative values. Near the tricritical temperature($T_0$), one usually expands $a$ as a power series of temperature difference - $a=a_0(T-T_0)$ where $a_0$ is a constant and $T$ is the temperature, and then searches for extrema in the free energy landscape. If all the coefficients of the free energy become positive then $F$ attains its minimum value at $\psi=0$. In other words, the system remains in the normal state. For $b<0$, however, the extrema of the free energy are located at those $\psi$-values which are solutions to the equation $a+b\psi^2+c\psi^4=0$, specifically the non-zero minima are located at $\psi^2=-b/2c\pm \sqrt{b^2-4ac}/2c$. Clearly a non-zero minimum exists if $b^2>4ac$. The minimum will be unstable energetically unless it satisfies $0=F(\psi=0)=F(\psi^2=-b/2c+\sqrt{b^2-4ac}/2c)$, which happens if $ac=3b^2/16$. So, in summary, in the free energy parameter space, we have an unstable finite-$\psi$ phase between $a=3b^2/16c$ and $a=b^2/4c$. The corresponding phase diagram is plotted in fig.(\ref{psi})
  
\begin{figure}
\centering
\includegraphics[scale=0.5]{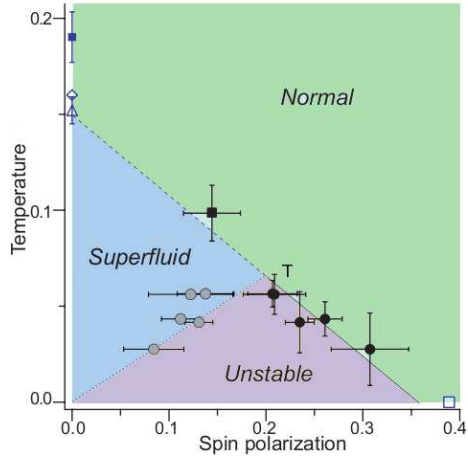}
\caption{Relevant portion of the phase diagram of spin-imbalanced, two-component Fermi gas, as reported by Shin et al(Nature,2008). The point 'T' denotes the tricritical point. The most important characteristics, which we are trying to point out in this paper, is that the width of the unstable region (shaded violet) increases with decreasing temperature, whereas the width of the unstable region, as obtained from a standard $\mathcal{O}(\psi^6)$ theory, decreases with decreasing temperature, as can be seen in fig.(\ref{psi}). }
\label{expic}
\end{figure}

Recently, an experiment by Shin et al\cite{shin08} captured the first-order transition very clearly by measuring spatial discontinuity in spin polarization . In the experimental phase diagram reported by Shin et al [see fig.(\ref{expic})], we note that the width of the unstable, finite-$\psi$ region \textit{increases with decreasing temperature}. It is, however, apparent that whatever polarization dependence we assign to the temperature-independent constant $b$ of the standard $\mathcal{O}(\psi^6)$ theory, the width of the unstable superfluid region \textit{decreases with decreasing temperature}. 

Keeping this discussion in mind, we set out to reconstruct the free energy of this imbalanced Fermi system following the work done on antagonistic order parameter in superconductor-ferromagnet phase transitions, several decades ago by Blount et al\cite{blount79}. Our phenomenological model free energy is a general Ginzburg-Landau type free energy with two order parameters, which are actually embedded in the system: one superconducting$(\psi)$, and another imbalance parameter$(m)$. This free energy explains the phase diagram properly and offers some valuable physical insights about imbalanced Fermi systems.

\section{The Model and Results}
\begin{figure}
\centering
\includegraphics{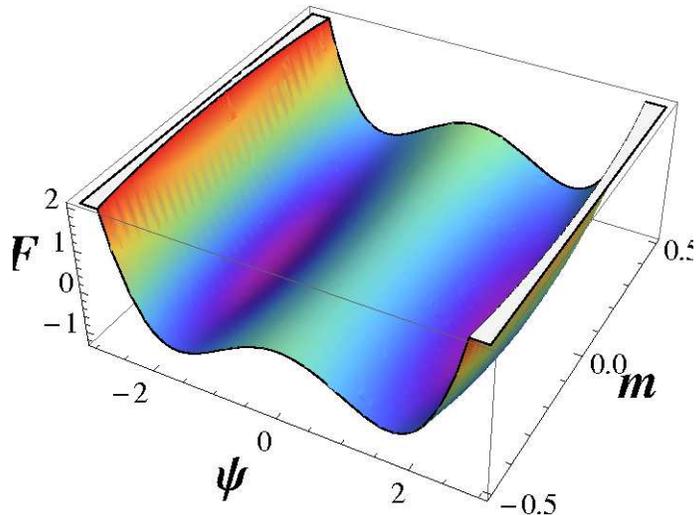}
\caption{This figure depicts our model GL free energy($F$) as a function of $\psi$ and  $m$. This clearly shows the double minima of $F$ with respect to $\psi$ and one minimum for $m$.}
\label{free}
\end{figure}

Our model free energy per unit volume is 
\beq
\label{f}
F= \frac{a}{2}\psi^2+\frac{b}{4}\psi^4+\frac{A}{2}m^2-mh+\frac{B}{2}m^2\psi^2
\eeq             
in which $\psi$ is superfluid order parameter, $a$ is given by $a=a_0(T-T_C)$ where $a_0$ is a constant and $T_C$ is the normal-superfluid transition temperature in absence of any imbalance. The parameter $b$ is a temperature independent constant. As usual in mean field theory, we have a uniform superfluid state for $T<T_C$ with $\psi^2=-a/b$. The imbalance parameter($m$) is the difference in the number of spin-up and spin-down fermions. This parameter represents magnetization in the superconducting scenario and polarization in the ultra-cold atomic gas. Following the standard Ginzburg-Landau formalism, we include a term proportional to $m^2$ in the free energy (where the coefficient $A$ is a constant) and the magnetic field (here chemical potential difference between the fermionic species) gives rise to an additional contribution $-mh$. We have neglected $\mathcal{O}(m^4)$ terms in the free energy as high imbalance destroys superfluidity. Finally, the $m^2\psi^2$ term, proposed after considering the symmetries of the system, signifies the interaction of the imbalance parameter with the superfluid order parameter. One has this freedom of choosing particular form of interaction while one is trying to explain the phase diagram of an imbalanced, two-species Fermi system at unitarity, because the particular functional form of the interaction does not matter in this resonantly-interacting regime. We have plotted the free energy in fig. (\ref{free}).    

To derive results for this model, we first minimise the Ginznurg-Landau free energy. Minimization results in two conditions on the order parameter - $\label{3}\bar{m}=h/(A+B\psi^2)$ and $\label{4}\psi(a+b\psi^2+B\bar{m}^2)=0$. Now for superfludity to be present in the system, one needs to solve the above equations simultaneously for non-zero solutions of the superfluid order parameter, and that is equivalent to solving
\beq
\label{2}
(a+b\psi^2)(A+B\psi^2)^2+Bh^2 = 0.
\eeq  
This result could also be viewed in a different way. We could integrate out the $m$-variable from Eq. (\ref{f}), leading to an effective free energy $F_{eff}$. Minimising this with respect to $\psi$ again leads to Eq. (\ref{2})\footnote{See appendix for detail calculations.}. This shows that results from conventional tricritical point can differ when $\psi^2$ is not small and our contention is to show that this is the point of difference from standard tricritical point.

From eq. (\ref{2}), we note that the superfluid order parameter vanishes \textit{continuously} along the curve $A^2a+Bh^2=0$. For temperatures greater than the tricritical temperature, a second order phase transition takes place as one crosses this curve to go from a normal($\psi=0$) to a superfluid phase($\psi\neq0$) or vice versa.

As our primary aim is to explain the phase diagram found by Shin et al, we will now check whether there exists a first order normal-superfluid transition in this system. If there exists a non-zero solution of the superfluid order parameter in the equation 
\beq
F(\psi,\bar{m})=F(\psi=0,\bar{m}|_{\psi=0}),
\eeq
 there will be a first-order phase transition in the system. We insert our particular free energy in this equation and found the following values for the superfluid order parameter 
\beq
\label{9}
\psi^2=-\frac{1}{3}\left(\frac{A}{B}+\frac{2a}{b}\right)+\frac{1}{3}\sqrt{\left(\frac{A}{B}+\frac{2a}{b}\right)^2-6\left(\frac{aA}{bB}+\frac{h^2}{bA}\right)}
\eeq

\begin{figure}
\centering
\includegraphics{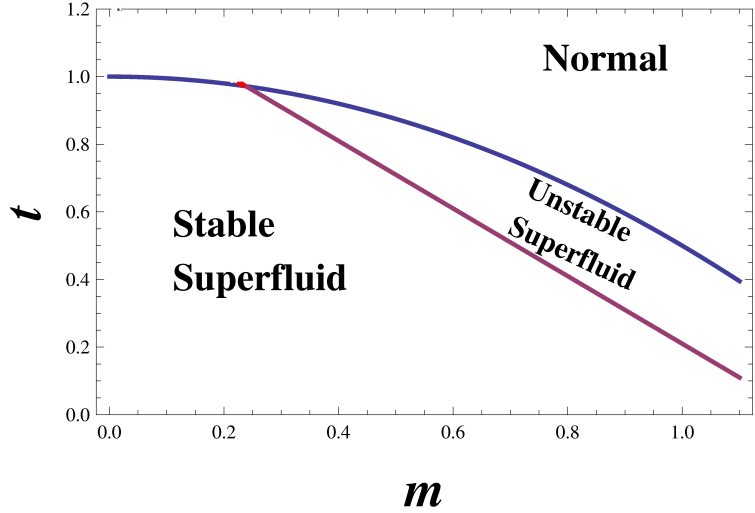}
\caption{Spin polarization($m$) vs. temperature($t=T/T_0$) phase diagram for a imbalanced two-component Fermi gas with resonant interactions. The area enclosed between the purple and blue line represents an energetically unstable region. The red dot is the tricritical point. For temperatures greater than the tricritical temperature superfluid-normal phase transition is second-order, while below it superfluid-normal phase transition is first-order which encounters an unstable phase en route. (Plotted using A=1.0, B=0.5 and b=0.2)}
\label{mp}
\end{figure}

Using the functional form of $\psi^2$, we get two equations which relate the magnetic field and critical temperature : $h^2=-aA^2/B$ and $h^2=A(bA-2aB)^2/8B^3$, across which a first-order normal-superfluid transition takes place. One of equations for the first order line is identical with the second-order line. The other curve is the purple line shown in fig(\ref{mp}).
The tricritical temperature($T_0$) can be easily determined by evaluating the intersection point of the two branches, which works out to be $a_0=A\left[  \left( \frac{b}{2B}-1\right) \pm\sqrt{1-\frac{b}{B}}\right] $. We designate this point as a tricritical point because it is the end point of a series of critical points, i.e., the order of the phase transition changes from second to first at this point.

Now let us figure out if there is any unstable region in the phase diagram within the scope of our model. This can be found out in this way- as we are treating this model at a mean field level, the phase of the superfluid order parameter is unimportant. Hence by demanding the square of the superfluid order parameter to be real, we find that there exist a unstable state for a range of temperature and magnetic field. According to the experimental data of Shin et al, in this region phase separation takes place. But from our free energy, which does not include any gradient term, we can only conclude that this region is unstable.

Another interesting feature of this model Ginzburg-Landau free energy is its  specific heat: we get a rather peculiar behaviour of this quantity near tricritical point. Generally in $\mathcal{O}(\psi^6)$ theories, where tricritical point is quite generic, at temperatures just below tricritical point specific heat diverges as $C\sim a^{-\frac{1}{2}}$. 

To calculate the specific heat of this our system, we differentiate twice the effective free energy ($F_{eff}$) with respect to $a$, which is nothing but scaled temperature, and we get
\begin{eqnarray}
\label{c}
C=\frac{\partial ^2 F_{eff}}{\partial a^2} &=&
\left( \frac{\partial \psi^2}{\partial a}\right)
+\frac{a}{2}\left( \frac{\partial ^2 \psi ^2}{\partial a^2}\right)
+\frac{b}{2}\left( \frac{\partial \psi^2}{\partial a}\right)^2
+\frac{b}{2}\psi ^2\left( \frac{\partial ^2 \psi ^2}{\partial a^2}\right)\nonumber\\
&-& \frac{B^2h^2}{\left(A+B\psi ^2 \right)^3 }\left( \frac{\partial \psi^2}{\partial a}\right)^2
+\frac{Bh^2}{2\left(A+B\psi ^2 \right)^2 }\left( \frac{\partial ^2 \psi ^2}{\partial a^2}\right) .
\end{eqnarray} 
So, as we can see from above equation, the nature of the specific heat is actually determined by the dependence of $\psi^2$ on $a$ which can be extracted from eq. (\ref{9}). There are two distinct and interesting cases where the specific heat behaves differently.  

Near transition point of the superfluid phase diagram, $\psi^2$ varies as $a^{1/2}$, in general, which can be derived from eq. (\ref{9}). If we put this $a$-dependence of $\psi^2$ in eq. (\ref{c}), we get $C=1+\frac{b}{2}+\frac{ba}{2}-\frac{B^2h^2}{(A+Ba)^3}$, which, near the transition point, becomes
\beq
C=1+\frac{b}{2}-\frac{B^2h^2}{A^3}.
\eeq    
So the specific heat have a jump discontinuity at the transition point.\textit{We emphasise, this is different from the standard tricritical behaviour of specific heat, which actually diverges as we approach the tricritical point from lower temperatures.}

Now let us consider the second case. Interestingly, we can expand our free energy to get a $\mathcal{O}(\psi^6)$ theory (for the steps see appendix). The coefficient of the $\psi^4$ term will be equal zero, which is required for a standard tricritical point, if $h_0^2=\frac{bA^3}{2B^2}$(from eq. (\ref{a3})). Now $\psi$ will be proportional to $a^{1/4}$, like in standard tricritical theory, if we set $\frac{2\mid a\mid}{b}=\frac{A}{B}+\epsilon$, where $\epsilon$ is a small number and in that case $\psi^2$ reduces to 
\beq
\psi^2=\frac{\epsilon}{3}+\frac{1}{3}\sqrt{\epsilon^2-6\left(\frac{h^2}{bA}-\frac{A^2}{2B^2}-\frac{\epsilon A}{2B} \right) }.
\eeq
Putting the value of $h_0^2$ in the above equation, we obtain $\psi^2=\frac{\epsilon}{3}+\frac{1}{3}\sqrt{\epsilon^2+\frac{3A}{B}\epsilon}$. As $\epsilon$ will tend to zero, $\psi^2$ will diverge as $\sqrt{\epsilon}$: finally we get the $\psi\sim a^{1/4}$ dependence. Doing a standard tricritical analysis we get $C\sim \left( a+\frac{Bh^2}{A^2}\right) ^{-1/2}$ , as expected. So indeed there is a point is the phase diagram, whose $(a,h)$ coordinates are $\left(-\frac{Bh^2}{A^2},\sqrt{\frac{bA^3}{2B^2}} \right)$, where we recover the standard tricritical behaviour.
      
Thus specific heat measurement will thus serve as an interesting check on whether the tricritical point is of the conventional type($\mathcal{O}(\psi^6)$ theory) or the findings over here. The main point of the last few paragraphs is that our model free energy encompasses the standard tricritical phenomena and, at the same time, points to some interesting physics which standard tricritical free energy does not account for. Before proceeding to a comparison of our free energy with the available microscopic calculation of this systems, we want to make one final comment about the special case in which we set $b=B$ in our free energy. In this case, the transition point of our model and the standard tricritical point becomes one, and one can have $C\sim a^{-\frac{3}{2}}$, as in standard tricritical case, also for the transition point of our model.  

To make the results quantitative, we connect the proposed free energy to the available microscopic calculations of the system\cite{maki64}.The microscopic calculation was done for a superconductor with an internal magnetic field generated by dilute magnetic impurity ions ,neglecting the orbital effects, which was a justified assumption in a two-dimensional superconductor. The magnetic field differentiates between spin-up and spin-down electrons and thus creates an imbalance. From Maki's calculation(Eq.(23) in ref(\cite{maki64}), we get that the coefficient of the fourth order term(in $\psi$) the free energy expansion is given by $-[(mp_0)/(8\pi^2(2\pi T)^2)]Re\sum_{n=0}^{\infty}(n+\frac{1}{2}+i\rho)^{-3}$ where $m,p_0$ are the mass and Fermi momentum of the electron, respectively, and $\rho=\mu H/2\pi T$. Near the tricritical point, we can approximate the above form to $-0.002mp_0$ after some algebra. In our theory, an expansion of the free energy in powers of $\psi^2$ gives the coefficient of the fourth order term to be $-(B^2/2A^3)(bA^3/2B^2-h^2)$ which can be approximated as $-(bB^2/2A^3)^\frac{1}{2}$ near the tricritical point. So we can put the following condition on the phenomenological coefficients of our theory $\left(\frac{bB^2}{2A^3}\right)^\frac{1}{2}=0.002mp_0$. We want to mention that as Maki's calculation was based on superconducting system, so this estimation is valid only in the deep-BCS limit. More recently, Sheehy\cite{sheehy09} had done a similar calculation.

Finally we mention one important point regarding this free energy. As our proposed free energy does not contain any gradient terms, it will not reproduce the Fulde-Ferrell-Larkin-Ovchinnikov(FFLO)\cite{ff,lo} superconducting state which is a superconducting state with a modulated order parameter. But one can extend this model free energy to include the gradient terms and study the FFLO state\cite{agterberg01,buzdin97,mora03}. Presently we are working in this direction.

\section{Conclusion}
To conclude, we have analysed the finite-temperature phase diagram of the two-component, imbalanced Fermi gas, as a function of population imbalance, using a phenomenological Ginzburg-Landau free energy. The novelty of our approach lies in using two order parameters, which, we assert, are actually embedded in the system: one superconducting and another imbalance order parameter. This approach is inspired by the Blount and Varma's Ginzburg-Landau free energy which was proposed to describe superconducting-ferromagnetic transition. Our analysis reproduces the phase diagram reported by experiment. We also point out that the specific heat behaves anomalously and provide a detailed analysis of the specific heat. We anticipate that the proposed  free energy, with two competing order parameters, will lead to improvements in our current understanding about the imbalanced Fermi gases.


\section*{Acknowledgement}
One of the authors (A.D.) thanks Council of Scientific and Industrial Research, India for financial support in the form of fellowship (File No. 09/575(0062)/2009-EMR-1).


\appendix
\section{}
If we integrate out magnetization from our model free energy, we get an effective free energy which is different from the standard $\mathcal{O}(\psi^6)$ theory:
\begin{eqnarray}
e^{-F_{eff}} &=& \int\mathcal{D}m e^{-F[m,\psi]} \nonumber\\
&=& \int\mathcal{D}m e^{-\left( \frac{a}{2}\psi^2+\frac{b}{4}\psi^4+\frac{A}{2}m^2-mh+\frac{B}{2}m^2\psi^2\right)}\nonumber\\
&=& const\times e^{-\left( \frac{a}{2}\psi^2+\frac{b}{4}\psi^4-\frac{h^2}{2\left( A+B\psi^2\right) }\right) }.
\end{eqnarray} 
So the effective free energy reads
\begin{eqnarray}
F_{eff}=\frac{a}{2}\psi^2+\frac{b}{4}\psi^4-\frac{h^2}{2\left( A+B\psi^2\right) }.
\end{eqnarray}
The interesting point is that we can recover the standard $\mathcal{O}(\psi^6)$ free energy from this effective free energy by expanding $F_{eff}$ near transition, where the value of $\psi$ is small:
\begin{eqnarray}
\label{a3}
F_{eff}=-\frac{h^2}{2 A}+\left(\frac{a}{2}+\frac{B h^2}{2 A^2}\right) \psi^2+\left(\frac{b}{4}-\frac{B^2 h^2}{2 A^3}\right) \psi^4+\frac{B^3 h^2 \psi^6}{2 A^4}+O[\psi]^8
\end{eqnarray} 
Now minimisation of the effective free energy with respect to $\psi$ can be done as follows:
\begin{eqnarray}
&&\frac{\partial F_{eff}}{\partial\psi} = 0\nonumber\\
&\Rightarrow &  \psi\left[ (a+b\psi^2)(A+B\psi^2)^2+Bh^2\right] = 0. 
\end{eqnarray}  
So, we have shown that the non-zero solutions of superfluid order parameter will be given by the solutions of eq. (\ref{2}).



\end{document}